\documentclass{emulateapj}

\usepackage{float}
\usepackage{subfigure}
\usepackage{verbatim}
\usepackage{color}

\begin{document}

\title{Evidence of residual Doppler shift on three pulsars, PSR B1259-63, 4U1627-67 and PSR J2051-0827}
%
%


\author
{Bi-Ping Gong$^{1}$, Z.Q. Yan$^{1}$}
\affil{$^{1}$School of Physics, Huazhong University of Science and Technology,
 Wuhan 430074, China}






\begin{abstract}
{ 
 The huge derivative of orbital period observed in binary pulsar PSR B1259-63, the torque reversal displaying on low mass X-ray binary, 4U1627-67 and the long term change of orbital period of PSR J2051-0827, seem  totally unrelated phenomena occurring at totally different pulsar systems.
In  this paper,  they are simply interpreted by the same mechanism, residual Doppler shift.  
In a binary  system with periodic signals sending to an observer,  the drift of the signal frequency actually  changes with  the  varying orbital velocity, projected to line of sight at different phases of orbit. 
And it has been taken for granted that the net red-shift and blue-shift of an full  orbit circle be   cancelled out, so that the   effect of Doppler shift to the signal in binary motion cannot be accumulated over the orbital period.  
However, taking the propagation time at each velocity state into account, the symmetry of the velocity distribution over the orbital phase is broken. Consequently, the net Doppler shift left in an orbit is non-zero. Understanding this Newtonian second Doppler effect not only  makes pulsars better laboratory in the test of gravitational effects, but also allows us to extract the angular momentum of the pulsar of PSR J2051-0827, $\leq 2\times 10^{43}gcm^2$; and  the accretion disc of 4U 1627-67, $7\times 10^{50} gcm^2/s$, respectively,  which are of importance in the study of structure of neutron stars and the physics of accretion disc of X-ray binaries. 
}
\end{abstract}
\keywords{stars: neutron---X-rays: binaries---X-rays: individual (PSR B1259-63, 4U1627-67, PSR J2051-0827, PSR 0823+26, PSR J0631+1036,   Her X-1, Cen X-3, GX 1+4, OAO 1657-415, Vela X-1, 4U 1907+09)}

\section{Introduction}

Doppler-shift is the change in frequency of a wave  or other periodic event  for an observer moving relative to its source, which has been applied to countless fields since its discovered in 1842. 
It occurs because the wave source has time to move by the time during previous waves encountering the observer. 




In a binary  system with periodic signal in orbital motion, the velocity of the signal project to the line of sight (LOS),  ${\bf v}$, changes with the orbital phase of the signal all the time. As a result, 
the  pulse of original frequency, $\nu_e$, is drifted with respect to an observer by, 
\begin{equation}\label{dnu}
 \Delta\nu=\frac{{\bf v}\cdot{\bf n}_p}{c}{\nu_e}=K{\nu_e}[\cos(\omega+f)+e\cos\omega ]
\end{equation}
where ${\bf n}_p$ is the unit vector of line of sight,  $c$ is the speed of light,    $K$ is semi-amplitude, $K=2\pi a_p\sin i/[cP_b(1-e^2)^{1/2}]$ (in which $a_p$ is the semi-major axis of the pulsar),  $e$ is he eccentricity of the orbit, $\omega$ is the advance of periastron,  and $f$ is the true anomaly, which is a function of time, $t$.

Because of such a orbital motion, the distance of the pulsar to the observer changes with $r=a_p(1-e^2)/(1+e\cos f)$, which corresponds to a time delay of, 
\begin{equation}\label{zc}
 \frac{z}{c}=\frac{r\sin i}{c}\sin (\omega+f)
\end{equation}
upon projecting to LOS, denoted by direction  $z$.  
Notice that the true anomaly, $f$, appeared in   Eq.$\ref{dnu}$ and  Eq.$\ref{zc}$ can be transformed to mean anomaly, $\bar{n}={2\pi}t/{P_b}$, where $t$ is the homogeneous time measured in the reference frame at rest to the center of pulsar. In fact, this time elapse can be measured as $t=nP$ (where $n=1,2,3...$, and $P$ is the spin period of the pulsar).

What is the net Doppler shift, e.g.,  to the pulse frequency of a binary pulsar system ?

For a given, $t$, we have orbital phase, $f(t)$, velocity ${V}(t)$ and hence $\Delta\nu(t)$. The integration to  $\Delta\nu(t)$ is $\int^{P_b}_{0}\Delta\nu(t)dt=0$, so that the blue and red-shift can be cancelled out in one orbit,  the Doppler effect of  Eq.$\ref{dnu}$  is thus  equivalent to the effect of Roemer delay of Eq.$\ref{zc}$. 
This conclusion has been taken for granted.

However such an equivalence needs to be reconsidered. 
Since pulsars are usually of distance of kpc to the Earth, the orbital phase of  a binary pulsar measured nowadays actually happens thousand years ago. As such a large time discrepancy produces a constant time delay, corresponding to the separation between the center mass of the binary system to the observer (after counting out out the proper motion),   which is negligible.


Whereas, the non-constant time delay is another story.  In a binary system, the true time measured by an observer is  
\begin{equation}\label{obs}
t_{obs}=t+\frac{z}{c}
\end{equation}
rather than the time $t$, describing  the orbital phase of the binary system.   
      
Substituting this $t_{obs}$ into Eq.$\ref{dnu}$ gives the true observational Doppler shift to the pulse signal. In other words, the orbital phase measured by an observer is  $f(t+z/c)$ instead of $f(t)$.



E.g., we have  two local times, $t_1$ and $t_2$,  defined  at rest to the center of the pulsar,  which determine the position  of the pulsar in orbital motion. If one takes the time, $t_k$ (where $k=1,2$),  as observer's time, then it  implies that the two pulses, sent at position 1 and 2, can reach the observer instantaneously. 

 

In such case, 
the two signals, which sent with a time discrepancy say, $t_2-t_1=1.0$s,  are thought to be detected as $\nu (t_1)$ and $\nu (t_1+1.0)$ respectively.

Whereas, the true situation is that the second signal is actually detected as, $\nu (t_1+1.0+\delta z/c)$, where $\delta z/c=z_2/c-z_1/c$. 


It is this additional time that changes the distribution of the projected velocity with respect to the observer. E.g., for a pulsar in circular motion around its companion star, it will have 1/4 orbital period of blue shift (close to), and 1/4 orbital period of red shift (away) at the two sides of the point of closest distance to the observer, without considering the propagation  time. But when the additional time is included, the turning point of the blue and red shift will change its position. And thus, the distribution of the projected velocity is no longer uniform with respect to the observer.

Consequently, net Doppler shift to the frequency of pulse is given\citep{Gong05},   
\begin{equation}\label{Gong05}
 \Delta{\nu}=\frac{1}{P_b}\int^{P_b}_{0}\Delta\nu(t)d(t+\frac{z}{c})=\frac{xK}{P_b}\nu\pi(1-\frac{e^2}{4})
\end{equation}
Notice that it is this additional time delay $dz/c$ that makes the shift of pulse frequency of Eq.$\ref{Gong05}$ non-zero. 
This result means that in every orbital period the signal frequency of one object in the binary system is shifted by a value described by Eq.$\ref{Gong05}$, which is a Newtonian second Doppler effect.   


Such a residual Doppler shift affects three typical binary pulsars, PSR B1259-63, 4U1627-67 and PSR J2051-0827 differently. In the following sections, they will be analysed one by one.

\section{PSR B1259-63}

PSR B1259-63 was discovered in a large-scale high-frequency survey of the Galactic plane (Johnston 1992). It is a radio pulsar in orbit about a massive, main-sequence B2e star. With an orbital period of 1237 d and companion mass of $>10M_{\odot}$,  the 23yr data\citep{Shan13} indicates a significant orbital period decay, $1.4(7)\times10^{-8}$, while the  timing residual versus orbital phase has similar shape as ten years ago\citep{Wang04}.

The huge derivative of orbital period corresponds to a time residual in one orbit, $\delta T_{o}=\dot{P}_b{P}_b\approx 1$(s). 

Can residual Doppler shift  reproduce such a significant timing residual ?
This can be tested easily by putting  $x=1296.3$s, $P_b=1236.7$d, $\nu=20.9 s^{-1}$ and $e=0.87$\citep{Wang04,Shan13}, into the residual Doppler shift in one orbit\citep{Gong05},  
\begin{equation}\label{dnu1259}
 \Delta{\nu}=\frac{xK}{P_b}\nu\pi(1-\frac{e^2}{4})=1.0\times 10^{-7} (s^{-1})
\end{equation}
Therefore, the predicted additional time residual in every  orbital period is  
$\delta T\approx\Delta\nu P_b/\nu\approx 0.6$s, which explains the residual,  $\delta T_{o}\approx 1$(s), corresponding to the significant derivative of orbital period\citep{Shan13}, with a flying color.


Further, the timing residual versus orbital phase 
can also be tested.  
Input $t$  into  Eq.$\ref{dnu}$, obtains phase, velocity, and hence frequency shift of pulsar at  moment $t$, 
from which the variation of $\Delta\nu(t)$ versus time is obtained as shown in  panel a of Fig. $\ref{f2}$.

As analysed above, the time $t$ is actually the local time at rest to the center mass of the pulsar. At this very time, the orbital phase of the pulsar is $f(t)$, whereas, the pulse signal sent at this phase is measured as $\Delta\nu(t+z/c)$ instead of $\Delta\nu(t)$.


The discrepancy between the two  shift of pulse frequency, $\Delta\nu(t+z/c)-\Delta\nu(t)$, is equivalent to O-C (observation-calculation). 
Expand  $\Delta\nu(t+z/c)$ in Taylor series, $\Delta\nu(t+z/c)=\Delta\nu(t)+\dot{\nu}(t)\cdot (z/c)+...$
Then the integration of the second term at right hand side gives,   
\begin{equation}
\label{pa} \int^{P_b}_0\dot{\nu}(t)\frac{z}{c}dt=\int^{P_b}_0\Delta\nu(t)\frac{z}{c}dt-\int^{P_b}_0\Delta\nu(t)\frac{\dot{z}}{c}dt
\end{equation}
Apparently, the first term at right hand side of Eq.$\ref{pa}$ equals zero, considering the trigonometric function of   Eq.$\ref{dnu}$ and Eq.$\ref{zc}$, and the second term is equivalent to Eq.$\ref{Gong05}$, which is actually  $\Delta\nu(t+z/c)-\Delta\nu(t)$. The variation of such a  discrepancy versus time is as shown in panel b of Fig. $\ref{f2}$.


The accumulated change of pulse frequency corresponds to a timing residual as shown in the bottom panel of  Fig. $\ref{f3}$, which is obtained by, 
\begin{equation}\label{acc}
 Res(t_k)= Res(t_{k-1})+\frac{\Delta\nu^{\prime}}{\nu}(t_{k}-t_{k-1})
\end{equation}
where $\Delta\nu^{\prime}={\Delta\nu({t_{k-1}+z_{k-1}/c})}-{\Delta\nu{(t_{k-1}})}$.


The curve with steps in the bottom panel of Fig. $\ref{f2}$  stems from accumulation of the Doppler residual at each orbit. The magnitude of each step actually corresponds to frequency shift given by Eq. $\ref{dnu1259}$. 

Cut off such a jump directly at the passage of precession of periastron,
we have timing residual versus time (orbital phase), as shown  in panel b of Fig. $\ref{f3}$.

As shown in Fig. $\ref{f3}$b, the dashed horizontal line corresponds to the level of 10 ms timing residual, which crosses each peak with a time scale of around 40 days. This well consistent with the facts that the emission is absent for nearly 40 days during the passage of periastron. Panel c displays the timing residual upon cut the part absent during the passage of periastron. 

Moreover, panel b of Fig. $\ref{f3}$b suggests that the closer the data to periastron, the larger the magnitude of the peak detected. Therefore, the reported glitch at MJD 50690.7\citep{Wang04} should be the epoch, which detects pulses  with shortest time interval with respect to the passage of periastron.       

Comparing the observational\citep{Wang04,Shan13} and simulated timing residual, as shown in panel a and c in Fig. $\ref{f3}$ respectively, there are still deviation between panel a and panel c, because panel a reduces the steps by assuming glitches\citep{Wang04}, which is not a constant in timing residual; while panel c is obtained by subtracting a  constant timing residual.







Consequently,  the strange timing behaviour of this pulsar  
 can be interpreted with available binary parameters\citep{Shan13}, and without introducing any additional parameters.

Notice that the amplitude of timing residual originating in Shapiro delay in a orbit is given,   $r=GM_c/c^3\sim 10^{-5}$s, which is much less than that of  the  residual Doppler effect of 1s in the case of  PSR B1259-63. 

If such a residual Doppler shift is evident in PSR B1259-63, which has a orbital period of years, what about a binary of orbital period as short as tens of minutes ?  

\section{4U 1627-67}

The accreting-powered pulsar 4U 1626-67 with a pulse period of 7.66 s,  was discovered by Uhuru\citep{Giacc72}. Although orbital motion has never been detected in the X-ray data, pulsed optical emission reprocessed on the surface of the secondary revealed, and thus confirmed  the 42 minutes orbital period\citep{Mid81}. 
This pulsar system is recognized as  a low mass X-ray binary (LMXB), with an extremely low mass companion of $0.04 M_{\odot}$ for i = 18deg\citep{Lev88}.

After  the steady spin-up observed during 1977 to 1989,  the torque reversal occurred during 1990 June, this pulsar began  steadily spinning down\citep{Chak98}. Interestingly, after about 18 yr of steadily spinning down, the accretion-powered pulsar 4U 1626-67 experienced a new torque reversal at the beginning of 2008\citep{CA10}.

The evolution of pulse frequency with two abrupt ``torque reversal", can  be understood in the context of  the 
 residual Doppler effect at long-term.



The residual Doppler shift of  Eq.$\ref{Gong05}$ predicts a  change of pulse frequency in an orbital period (with $e\approx 0$),  
\begin{equation}\label{dotnu}
 {\Delta\nu}=\frac{2\pi^2a_p^2\nu}{P_b^2c^2}\sin^2i
\end{equation}
The  observations\citep{CA10} corresponds to the change of $\delta\nu$ in dozens of years, containing numerous orbital period, $P_b=42$min. Such a long-term variation can be obtained by  the  integration of  $\sin^2 i$ of Eq.$\ref{dotnu}$ by time.    

\begin{figure}
\includegraphics[width=0.45\textwidth]{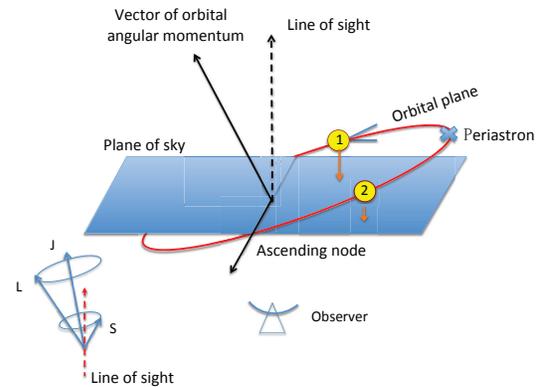}
\caption{\small Pulse signal sent by a pulsar in orbital motion. Pulses sent at different orbital phases correspond to different propagation time to reach the observer. The bottom left panel represents the 
spin-orbit coupling process, where the orbital and spin angular momentum, $L$ and $S$,  precess around the total vector $J$,  with  $L$ and $S$ are at opposite side of $J$ instantaneously.
Therefore,  the misalignment angle between $S$ and  LOS varies with the precession. 
\label{f1}}
\end{figure}

\begin{figure}
\includegraphics[width=0.45\textwidth]{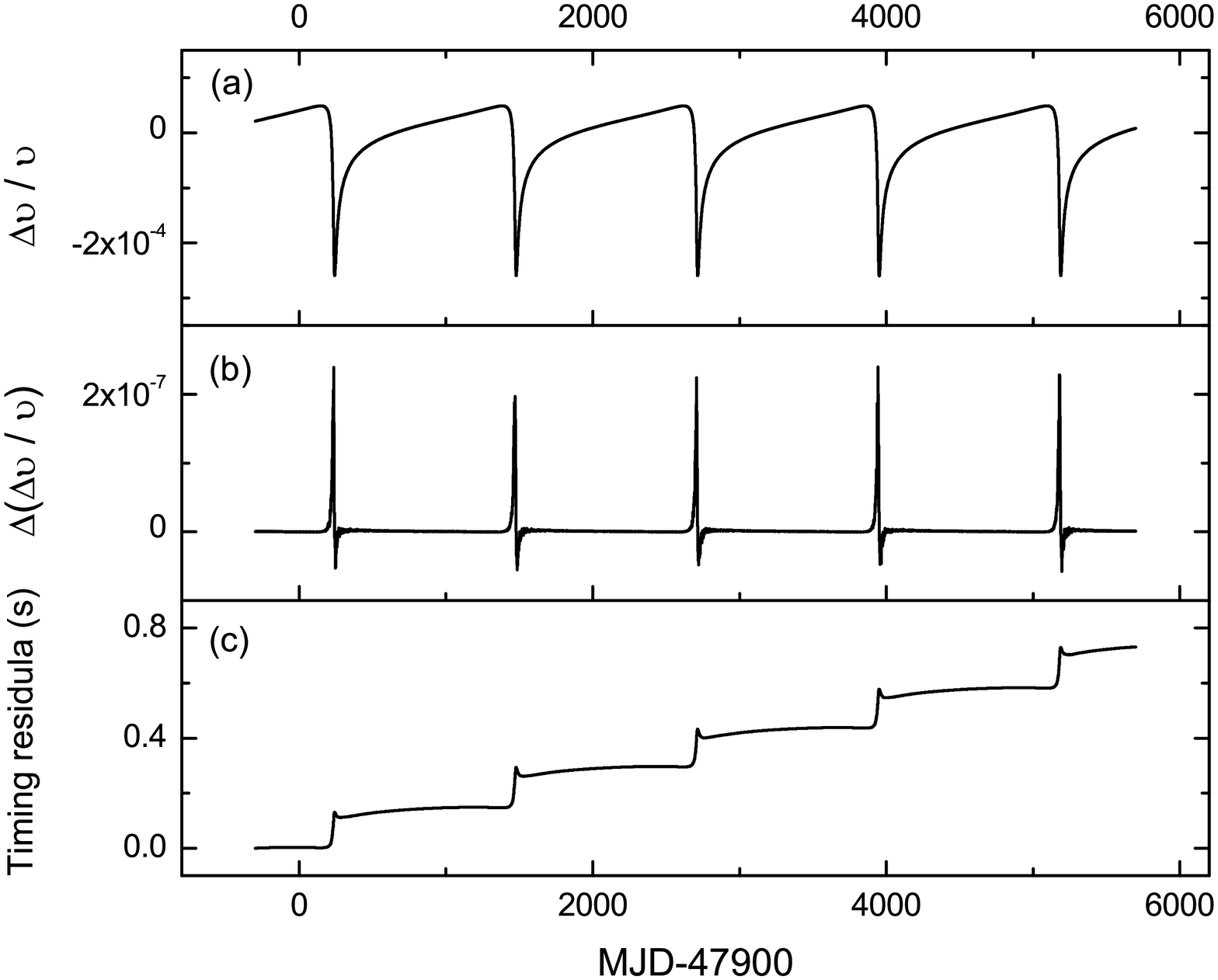}
\caption{\small Simulation of the normal Doppler  and residual Doppler on PSR B1259-63.
(a) Input $t$  into  Eq.$\ref{dnu}$, obtains Doppler-shift to the pulse signal at different time. (b) Shows the discrepancy between Doppler-shift with and without propagation time, $\Delta\nu(t+z/c)-\Delta\nu(t)$. (c) Timing residual originated in accumulated Doppler shift, represented by the curve with steps.
\label{f2}}
\end{figure}

\begin{figure}
\includegraphics[width=0.45\textwidth]{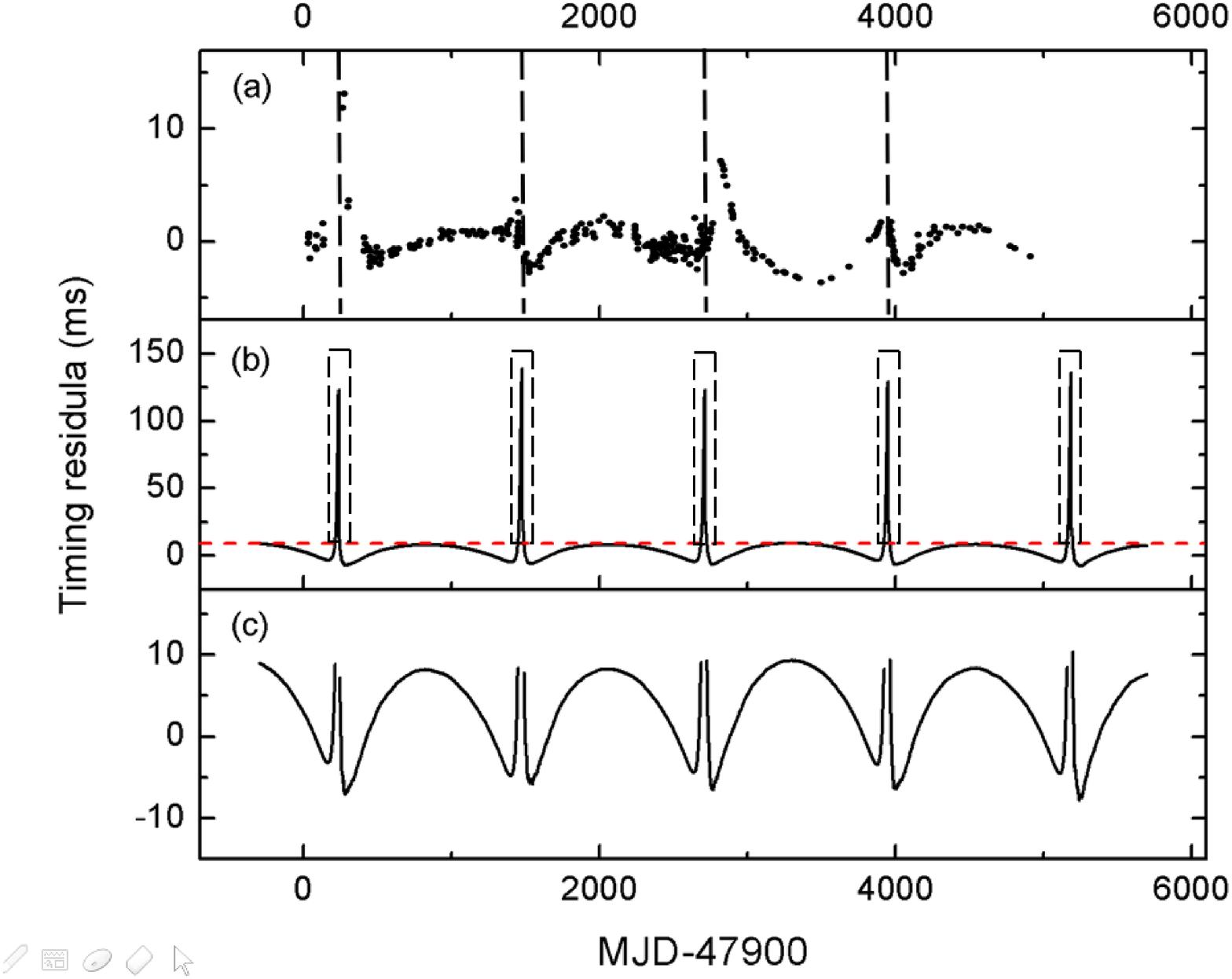}
\caption{\small Comparison between observed and simulated timing residual versus time of PSR B1259-63. 
The top panel is the observed timing residual of\citep{Wang04}. The middle panel  displays the result after  counting out the steps in accumulated Doppler shift, as shown in panel c of  Fig.$\ref{f2}$, 
which is obtained by cut off  two neighbouring steps at exactly the periastron. 
The dashed horizontal line corresponds to the level of 10 ms timing residual, which crosses each peak with a time scale of around 40 days. This well consistent with the facts that the emission is absent for nearly 40 days during the passage of periastron. 
\label{f3}}
\end{figure}

The  spin-orbit coupling of binary system is likely responsible for the variation of $i$, as shown in the bottom left panel of Fig. $\ref{f1}$ .

Under such a coupling effect, both the spin and orbital angular momentum precess around the total angular momentum.  
The precession of the spin angular momentum, which is so called geodesic precession,  results in the change of pulse profile which have been observed  in a number of  pulsar binaries, e.g., PSR 1913+16\citep{Weisberg02,Kon03}.

And  the variation of the orientation of the orbital plane, represented by  the angular momentum of the orbit, $L$, leads to evolution of the orbital inclination angle, $i$,  denoting the misalignment angle between $L$ and  LOS, as shown in the bottom left of Fig.$\ref{f1}$.

The time scale of such a spin-orbit coupling depends on  the orbital period.  E.g., for PSRJ0737-3039 of orbital period of 2.4 hours, the long-term spin-orbit coupling effect is about 70 years, and for PSR 1913+16 with orbital period of 7.7 hours, the time scale is 300 years. The precession of orbital plane of 4U1627-67 is calculated\citep{BO75},
\begin{equation}\label{BO75} 
 \Omega=\frac{GS(4+3M_c/M_p)}{2c^2a^3(1-e^2)^{3/2}}\hat{S}
\end{equation}
where $S$ and $\hat{S}$ are the magnitude and the unit vector of the spin angular momentum around the pulsar respectively.

As shown by Eq. $\ref{dotnu}$, the magnitude  of $\delta{\nu}$ is determined by $P_b$, $\sin i$, and $a_p$, in which $P_b$ is given by observation directly,   while $\sin i$ and $a_p\equiv aM_c\sin i/(Mc)$ (where $M=M_p+M_c$) are also constrained by observations\citep{Mid81,Lev88}.

We can make $M_c$ and  $i$ as free parameters, and use Eq. $\ref{dotnu}$ to fit 
 both the amplitude and time scale of the observed change of pulse frequency\citep{CA10}.  

Obviously, the variation of $\delta\nu$ is due the change of $i$, which is in turn originated in the S-L coupling as shown in the bottom left of Fig. $\ref{f1}$. As the time scale of variation of $i$ is determined by Eq. $\ref{BO75}$,
this set constraint not only on the companion mass, but also on the  spin angular momentum around the pulsar, $S$. 

with $P_b=42$min and $M_p=1.4M_{\odot}$, we find that the best orbital inclination  (average), companion mass and spin angular momentum are $i=0.3$rad, $M_c=0.16M_{\odot}$, and $S= 6.9\times 10^{50} gcm^2/s$, respectively as shown in Table~1. 



Such a spin angular momentum  is much larger than that of NS and WD, which   suggests that it stems from accretion disc around the pulsar rather than the pulsar itself.  The existence of such a disc is supported by the X-ray emission lines\citep{CA12}.









\begin{figure}
\includegraphics[width=0.45\textwidth]{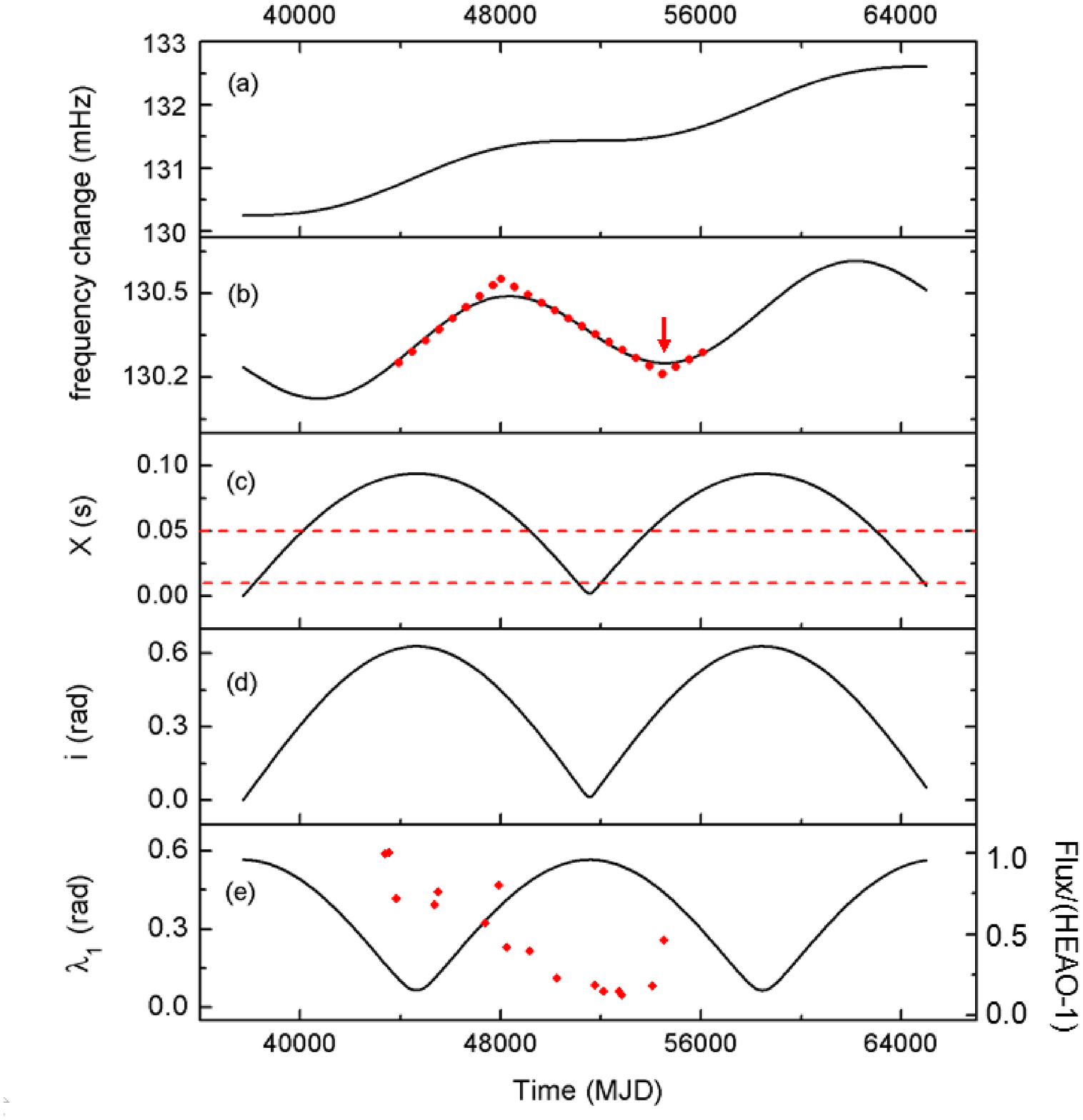}
\caption{\small Fitting results of 4U 1627-67. (a) The resultant integration of Eq.$\ref{dotnu}$ by time. (b)  The change of $\delta\nu$ after the constant increase of  $\delta\nu$ is cancelled by the spin-down of pulsar spin of $\dot{\nu}=-8.9\times 10^{-10}$Hz/s. The dots correspond to observational data. (c) Predicted evolution of the projected semi-major axis of the pulsar, $x$. The dashed horizontal line at the bottom corresponds to the level of $x=10$ms, the MJD corresponds to such levels can be found. And upper dashed horizontal line corresponds to the level of  $x=50$ms.  (d) Predicted evolution of orbital inclination, $i$, the variation of which is determined by the effect of S-L coupling. (e) Predicted evolution of the misalignment angle between spin angular momentum of the accretion disc and the LOS. The dots correspond to observational flux change\citep{CA10}.   
\label{f4}}
\end{figure}

With the fitting parameters of Table~1, the resultant integration of Eq.$\ref{dotnu}$ by time is shown  at the top panel of Fig. $\ref{f4}$,   
which exhibits a constant increase of  pulse frequency of   $\dot{\nu}=8.9\times10^{-10}$Hz/s,  due to  Eq.$\ref{dotnu}$ is always positive. 




The constant increase of pulse frequency  have been  cancelled by the spin-down of pulsar spin.
However, the cancellation is not perfect, the positive (residual Doppler shift) overwhelms the negative (spin-down) a little, so that  the frequency variation vs time has a overall trend of increase, as shown in panel b of Fig. $\ref{f4}$.


Such an increase of $\delta\nu$ predicts that next peak of $\delta\nu$ vs time, immediately after that of 2000 (MJD 54500),  should occur around MJD 62140, and the amplitude of which must be higher than that of MJD 54500.
This prediction can be tested soon.  


On the other hand, the  panel d of Fig. $\ref{f4}$  shows the spin-orbit coupling induced  variation of $i$, 
which reaches its maximum and minimum around MJD 44600 and MJD 51500 respectively. 
Accordingly, $x\equiv a_p\sin i/c$, varies as shown in panel c of Fig. $\ref{f4}$, and the discrepancy of which  can be up to 10 times in magnitude.  
This  explains different $x$  measured by different authors at different times,   
  $x=0.36\pm0.10$s\citep{Mid81,Lev88} and $x=8ms-3ms$\citep{Chak97}.

The maximum value of $x$ should appear again around MJD 58450, as shown in panel c of Fig. $\ref{f4}$, which is 5 years after Jan 1, 2014 (of MJD 56658), this prediction will test the model from another aspect.








The spin-orbit coupling process is actually the precession of both vectors, $L$ and $S$ around the total vector $J$,  with  $L$ and $S$ are at opposite side of $J$ instantaneously, as shown in the bottom left panel of Fig. $\ref{f1}$. 
Therefore,  their misalignment angle with LOS varies differently with the precession, one misalignment angle at maximum corresponds to the minimum of the other.  

 E.g., the misalignment angle between the spin angular momentum, S and LOS reached the minimum at around MJD 44630, as shown in  panel e of Fig. $\ref{f4}$. And since the spin angular momentum vector usually aligns with the outflow of a X-ray binary system, the minimum misalignment angle corresponds to the strongest effect of Doppler boosting, which automatically explains the  flares of 4U1627-67 in early 1980s (of MJD around 44630). 
 
Later, the increase of such a misalignment  angle weakens the Doppler boosting,  and hence prevents the flares from being observed afterword. 




According to panel e of Fig. $\ref{f4}$, in around MJD 58460, the misalignment angle will return to the level like the early 1980s again. In other words,  this predicts that the flaring stage like the early 1980s will happen in 2019.   

By the prediction of the bottom panel of Fig. $\ref{f4}$, the misalignment angle starts increasing at MJD 44630, till the turning point around MJD 51570. This predicts  a flux decrease during MJD 44630-51570. And the flux will start increasing after MJD 51570, as indicted by the bottom panel of Fig. $\ref{f4}$. 
This is well consistent with the X-ray light-curve observed\citep{CA10}, which shows that the enhancement of flux density of 4U1627-67 occurs surely before MJD 54000.    

Moreover, as shown in panel b of Fig. $\ref{f4}$, the turning point of $\delta\nu$ is between MJD 54000-56000, this again  is consistent with the observed evolution of $\delta\nu$\citep{CA10}, in which the turning point is surely after MJD 54000. 

Therefore, the new model not only explains the amplitude and time scale of $\delta\nu$, $x$, and flares of 4U1627-67, 

but also their turning points and correlation. All of these are interpreted by a simple scenario of the binary system, the effect of residual Doppler shift. It is more difficult to understand if all these happen by chance. 
The future correlation of these three parameters, are clearly predicted in Fig. $\ref{f4}$, some of which can be further tested soon.





Further more, 
the spin angular momentum of the accretion disc around the  pulsar inferred by the residual Doppler shift is of importance in the understanding  of the timing and emission  of X-ray binaries.




Moreover,  two radio pulsars, PSR 0823+26 and PSR J0631+1036 exhibit an abrupt change of timing residual\citep{Baykal99,Yuan10}. And other X-ray pulsars,
Her X-1, Cen X-3, GX 1+4, OAO 1657-415, Vela X-1\citep{B97}, and 4U 1907+09\citep{I09}, exhibit 
torque reversal as  4U 1627-67, 
which 
suggest that they may stem from the same  mechanism as 4U 1627-67 does.


\section{PSR J2051-0827}
PSR J2051-0827 is the second eclipsing millisecond pulsar system.  Long-term timing observations have shown secular variations of the projected semi-major axis and the orbital period of the system. These two variations has been interpreted separately. The variation of the former has been interpreted  as S-L coupling in the binary system, whereas, and the change in the latter is explained by tidal dissipation leading to variation in the gravitational quadrupole moment of the companion\citep{Laz11}.      


The residual Doppler effect provides simple and unified mechanism that interprets both variations without introducing any additional parameter to this binary system.


We first analyses the magnitude of variation of the two values, $x$ and $P_b$. 

From the observation\citep{Laz11}, the variation amplitude of $\Delta x$ in about 2000 days is of $10^{-4}$. This corresponds to a variation of $\sin i$  of  $10^{-4}$ in the same time interval,  since the change of $i$ is the most probable origin of the variation  of $x$, as shown in Fig.$\ref{f1}$.  

The usual recognized  the companion mass of $0.05M_{\odot}$\citep{Laz11} is actually difficult to satisfy   $x=0.045$lt-s and $i=40^{\circ}$ simultaneously. It is found that  a companion mass of $M_{c}=1.0M_{\odot}$ not only satisfies the observational constraint, $x=0.045$lt-s, in  the case of $i=33^{\circ}$, but also explains the variation of both $x$ and $P_b$.

By the observed binary parameters\citep{Laz11}, $P_b=0.1$day, and assuming $M_p=1.4M_{\odot}$ and $M_{c}=1.0M_{\odot}$, we have semi-major axis of the binary, $a=5.9\times 10^{9}$cm, through Kepler's third law.
And putting these parameters into Eq.$\ref{dotnu}$ obtains $\delta\nu=4.5\times 10^{-7}\sin^2i$(Hz), which corresponds to an additional time delay of,  
$$\delta T=\delta\nu P_b/\nu=1.8\times10^{-5}\sin^2 i \,, (s)$$ 
in each orbital period. Dividing such a $\delta T$ by $P_b$, we have the  first derivative of  the orbital period, 
$\dot{P}_b=0.2\times 10^{-8}\sin^2 i$.

Expanding it in Taylor series, 
 $$\dot{P}_b=0.2\times 10^{-8} [\sin^2 i_0+\Delta i\sin 2i_0+O(\Delta i^2)]$$ 
The first term at the right hand side of above equation is a constant, predicting  $\dot{P}_b\approx 6\times 10^{-10}$ in the case $i_0=33^{\circ}$.  
And the second term actually predicts a variable  $\dot{P}_b$ of $\sim 10^{-13}$ (with $\Delta i=10^{-4}$).

If the constant part of  $\dot{P}_b$ (first) is completely absorbed by spin parameters like $\nu$ and $\dot{\nu}$, then one would observe a varying $\dot{P}_b$ with equal amplitude of variation, as given by the second term in the equation above. However, down trend in the top panel of Fig.5\citep{Laz11}  suggests that the constant part of $\dot{P}_b$ is not completely removed. The net value of the two terms is $\dot{P}_b\sim 10^{-12}$.    

Consequently, at the time interval of T=1000 days ($1\times 10^4$ orbits), the change of orbital period is   $\delta P_b=\dot{P}_b T\sim 1\times 10^{-8}$, which  consists with  the observational change of $P_b$\citep{Laz11}.



In fact, the variation of  $\sin i$  at the level  $10^{-4}$\citep{Laz11} implies a ratio between spin and orbital angular momentum of $S/L\sim 10^{-4}$ (assuming a right angle between S and L).


The orbital angular momentum is $L=\mu\bar{\omega} a^2(1-e^2)^{1/2}\approx 4.5\times 10^{49} gcm^2s^{-1}$ (where $\bar{\omega}=2\pi/P_b$). 
This predicts a minimum
spin angular momentum of $4.5\times 10^{45}gcm^2s^{-1}$. The means the minimum moment of inertia of the pulsar is about $2\times 10^{43}gcm^2$.


Moreover, as shown in Fig.$\ref{f1}$, $S/L\sim 10^{-4}$ requires that 
the misalignment angle between the L and J is of order $10^{-4}$, and  the precession scenario is that both L and S precesses about J rapidly.   

Then precession velocity of the orbital plane can be treated as\citep{BO75,Apo94,Wex99}, 
$$\Omega=\frac{3G\bar{\omega}(M_c+\mu/3)}{2c^2a(1-e^2)}=3.9\times 10^{-8}s^{-1}$$, 
which well consistent with that required by observation, $2\pi/(2.7\times 10^{3}\times 86400)=3\times 10^{-8}s^{-1}$.       

In summary, at the expanse of increase companion mass to $1.0M_{\odot}$ (without introducing any new parameter), both the amplitude and time scale of the variation of $x$ and $P_b$ are interpreted.   

\section{Discussion}







Such a  residual Doppler shift should exist in all binary pulsars, 
and the timing residual produced by it is  much larger than that of Shapiro delay in each orbital period.  And differing from  the effect of Shapiro delay, residual Doppler shift can accumulate over time,  which results in puzzling timing behaviours at the time scale much longer than the orbital period. 

Where does it go in ordinary binary pulsars ? The constant part of the effect can be absorbed by either $\nu$ or $\dot{\nu}$ (or both). E.g., for PSR B1259-63, its residual Doppler shift given by Eq.$\ref{dnu1259}$ divided by the orbital period yields,
 $\dot{\nu}=1\times10^{-15}$Hz/s, which is much less than the amplitude of its spin down, $\dot{\nu}=-1\times 10^{-11}$Hz/s observed\citep{Shan13}.
Therefore,  its contribute to  $\dot{\nu}$ is not easy to figure out from  the observed one. It is the large amplitude of variation of $\delta\nu$ at each periastron passage  that makes it different (which is a short-term effect, at the time scale of the orbital period). 



And for  4U 1627-67, the residual Doppler shift induced timing residual makes the pulse frequency change steadily for 18yr (for ordinary binary pulsars the time scale is at least one order magnitude larger), which can be hidden in the parameter, e.g., $\dot{\nu}$, before the sign changes that  reveal its identity.      

As a result, most of the binary pulsar system, with binary period not so large and so less,  would not be so lucky as the three pulsar systems that display so distinguished timing properties so that the new effect can be extracted.    

Thus, understanding residual Doppler shift sheds new light on pulsar  timing. 
On the other hand,  the  counting out this effect is 
of importance not only in the test of  effects of general relativity like Shapiro delay, Einstein delay, relativistic precession, but also  in pulsar timing array  using pulsar timing to detect  gravitational wave. 

In addition, obtaining the spin angular momentum of the pulsar and the accretion disc around the compact object are also of importance in the understanding the nature of both pulsar and X-ray binaries.

\begin{table}
\begin{center}
\caption{\bf Different fitting parameters for  4U 1627-67. }
\begin{tabular}{cccccc}
\hline
\hline
 & $M_S$ & $M_c(M_{\odot})$ & $\lambda$(deg) & $\lambda_{LJ}$  & $S$($1\times 10^{50}gcm^2$) \\
a & 1.9 & 0.126 & 18.0 & 34.4  & 9.2 \\
b & 1.9 & 0.251 & 14.4 &  18.0  &  9.3 \\
\hline \hline
\end{tabular}
\end{center}
\end{table}

\acknowledgments
We thank Manchester, R.N., Ni, W.T., Wang,N., Peng, Q.H., Zou, Y.C., Yuan, J.P., and Zhang, M.   for  helpful 
discussion. This research is supported by the
National Natural Science Foundation of China, under grand
NSFC11373018, and Beyond the Horizons 2012.

{}

\end{document}